\documentclass[aps,prl,reprint]{revtex4-1}
\usepackage{graphicx}
\usepackage{bm}
\usepackage{epsf}

\begin{document}

\title{ Onset and saturation of ion heating by odd-parity rotating-magnetic-fields  in a field-reversed configuration}


\author{Alexandra S. Landsman} 
\email{alexandra.landsman@phys.ethz.ch}
\affiliation{Department of Physics, ETH Zurich}
\author{Samuel Cohen}
\affiliation{Princeton Plasma Physics Lab}
\author{Alan Glasser}
\affiliation{Los Alamos National Lab} 
 
 \maketitle

Heating of  figure-8 ions by odd-parity rotating magnetic fields ($RMF_o$) applied to an elongated field-reversed configuration (FRC) is investigated.  The largest energy gain  occurs at resonances ($s \equiv \omega_R/ \omega$) of the $RMF_o$ frequency, $\omega_R$, with the figure-8 orbital frequency, $\omega$, and is proportional to $s^2$ for $s-even$ resonances and to $s$ for  $s-odd$ resonances.    The threshold for the transition  from regular to stochastic orbits  explains both the onset and saturation of heating. The FRC magnetic geometry lowers the threshold for heating below that in the tokamak by an order of magnitude.

Heating, {\it i.e.,} stochastic energy gain, of charged particles by time-varying fields is a  complex and fundamental phenomenon  critically important to  as diverse areas of plasma physics as fusion research\cite{Stix:92} and plasma processing.  Well known are the effects of simple resonances and  particle collisionality on the heating of magnetized plasmas. Far less well explored is the role of an inhomogeneous static magnetic-field geometry.   Because of its relevance to space plasmas\cite{Chen:92}, plasma processing\cite{Rax:00}, and magnetic-confinement controlled-fusion research\cite{Tuszewski:88}, the field-reversed configuration (FRC, see Fig. 1) -- with its poloidal field nulls, lack of toroidal field, and strong field gradients -- is an important system in which to explore the effects of magnetic field geometry on particle dynamics under the influence of time-varying fields. 

Even with axial symmetry, a static FRC allows charged-particle orbits that are regular or ergodic\cite{Landsman:03}. First studies of single-particle orbits in FRCs assumed time invariance and spatial symmetries  that reduced the problem to one or two dimensions, allowing Kolmogorov-Arnold-Mosher (KAM) surfaces to exist\cite {Lichtenberg:92} and limiting excursions in phase space. The addition of a  rotating magnetic field (RMF)\cite{Landsman:03} breaks the angular invariance of the FRC, creating a three-dimensional system without bounding KAM surfaces and opening the possibility for large excursions in phase space and energy.  These excursions can have beneficial results, such as ion heating\cite{Cohen:00}, or detrimental ones, such as loss of confinement. In this paper we present studies of ion orbits in  FRCs with RMF applied: the goal is to understand the threshold for chaos and the role of resonances in the non-linear growth and subsequent saturation of ion energy.  We restrict attention to the novel  odd-parity RMFs ($RMF_o$)  because of field closure and encouraging recent experimental results. We show that the same mechanism is responsible for the initial ion heating and its ultimate saturation.

Studies of stochastic ion heating  by perpendicularly propagating electrostatic  waves in tokamaks were performed with similar Hamiltonian techniques and research goals. The results we report are markedly different  because of fundamental differences in the magnetic field geometry of the two devices.

Earlier papers \cite{Cohen:00}, which used the RMF numerical code to investigate  $RMF_o$s applied to FRCs, showed that the relevant frequency range for ion heating was  broad, $|\Omega| \sim 0.2-2 $, where  $\Omega  \equiv  {\omega_{R}/  \omega_{ci} }$, $\omega_R$ is the $RMF_o$ frequency,  $ \omega_{ci}= q B_a/ m c$ is the ion-cyclotron frequency in the  axial field at the FRC's center, $B_a$, $m$ is the ion mass, and $q$ is the ion charge. These papers reported significant ion heating  even for low relative RMF amplitude, $B_R$:  $B_R/B_a \sim 5 \times 10^{-4}$.  Phase de-coherence of ion orbits, with respect to the periodic electric fields created by the $RMF_o$, is a necessary condition for ion heating.  Strong gradients and regions of field reversal in the FRC  provide locations for possible phase de-coherence. For a 10-cm FRC having an ion density of $10^{14}$ cm$^{-3}$ and an ion energy of 100 eV, Coulomb collisions will be $10 \times$ less frequent than the stochastic effects described herein\cite{Cohen:00}.

The question arose whether, in spite of the existence of strong field  gradients,  ion-cyclotron resonances (ICRs) were important to ion heating. We show that  ICRs are important, but with significant differences from the standard ICR picture.  More rapid heating occurs at low $B_R/B_a$ for figure-8 orbits (see Fig. 1b)) than  for cyclotron orbits, though the latter  have a more clearly resonant interaction with $RMF_o$.     Figure-8 orbits cross the field-reversal and strong-gradient regions (twice) every orbit cycle,  possibly losing phase coherence at each traversal.   In contrast, cyclotron orbits may only incur phase de-coherence at the less frequent excursions to the axial extremes of their orbits. Betatron orbits have a less non-linear nature and hence are also less well heated than figure-8 orbits.  Because figure-8 orbits are  representative of a large fraction of ions in hot fusion FRC plasmas and because they represent the physically interesting situation of motion in a double potential well\cite{Neishtadt:97}, we focus on them.  The studies presented herein also clarify why high-energy  orbits  tend to interact regularly with $RMF_o$, leading, importantly, to a saturation of ion heating by $RMF_o$ and a method for tuning ion energy.   

We follow Ref.  \cite{Cohen:00} by using  the same equations for the  $RMF_o$ and a Solov'ev equilibrium for the FRC, with the notation, see Fig. 1a): $R$ = FRC separatrix radius in midplane; $Z$ = FRC axial half length; $\kappa \equiv Z/R$, FRC elongation; $r$ = radial coordinate;  $z$ = axial coordinate;  $\phi$ = azimuthal coordinate;  $p_i$  =  canonical momenta; $P \equiv 2 p_\phi/q B_a R^2$, normalized $p_{\phi}$; $\bf A$ = vector potential of RMF and FRC;   $\psi = \phi - \omega_R t$; $k=l \pi  /\kappa R$ = axial wave number of the $RMF_o$;  $l$ = axial mode number; and time, $\tau$, in units of $2\pi/\omega_{ci}$.  

The shape of the effective potential-energy surface on which an ion moves depends on $P$ and $z$\cite{landsman03, landsman07}:  figure-8 orbits may be confined to the $z=0$ subspace, or  to a potential-well minimum above or below $z = 0$, or may oscillate across $z=0$.  Orbits confined to the $z=0$ subspace  are amenable to an analytic analysis and are the appropriate choice to analyze because of the  $RMF_o$'s electric fields,  $ \mathcal{E}_r$ and $ \mathcal{E}_{\phi}$, there. 
Each cross-section in $z$ is either a double potential well, allowing both cyclotron and figure-8 orbits, or a raised potential, corresponding to betatron orbits.   Cyclotron orbits feel a force towards larger $|z|$, thus eventually enter a region where the barrier between the double wells is low enough for them to traverse, thereby becoming figure-8 orbits.    Since cyclotron orbits interact regularly with RMF, except at these  axial extremes,  their random fluctuations in energy appear less frequently than for orbits which are always figure-8.  Moreover, figure-8 orbits have greater radial  excursions, hence gain more energy from the radial electric field of the $RMF_o$. It follows that the heating of figure-8 orbits is an upper limit for the heating of all ions in the FRC and that the threshold for heating is highest in the $z=0$ subspace. Extensive numerical simulations, performed with the $RMF$ code, confirmed this.   

We first examined whether the broad $\Omega$ range for heating is due to resonances at the fundamental ICR frequency.  As described below, the answer is no. Instead high-harmonic resonances occur because the frequency of the figure-8 orbit is highly nonlinear. (High-harmonic resonances have recently been observed in an RMF experiment.)  As the energy of a figure-8 orbit decreases, the ratio $s \equiv \omega_R/\omega$ increases because the ion's frequency, $\omega$, slows down as it gets closer to the phase-space separatrix created by the hump in the double potential well.  A set of resonances with the $RMF_o$  occurs at integral values of $s$.

 Figure 2a) shows the $RMF$-code-calculated time dependence of ion energy for two values of $B_R$ for a $1$-keV ion initiated in a figure-8 orbit with zero axial velocity in the $ z=0$ subspace of an FRC having $R = 10$ cm, $\kappa = 5$,  $\Omega = 0.9$,   and $B_a = 20$ kG. Regular motion, with a clear $s=5$ component, is seen for $B_R = 2$ G. The energy fluctuations are  small ($\sim15 \%$) for $B_R = 2$ G and large, $> 100 \%$, for $B_R =20$ G. For $B_R = 2$ G, the fast Fourier transform (FFT) of ion energy, Figure  2 b), shows sharp peaks in frequency space, indicative of regular motion.  The separation between peaks, $\Delta f$, is $0.207 \pm 0.002$, in units of $ \omega_{ci}$. For $B_R = 20$ G, the FFT shows broadband noise at a $~ 30 \times$ higher absolute level.  Under these conditions, betatron (also shown in Fig. 2b)) and cyclotron orbits (not shown) display regular motion --- sharp peaks in their FFTs --- with energy fluctuations less than $7 \%$, even for $B_R = 20 \ G$. Figure 2c) shows $\Delta f$ {\it vs} $\tilde E$, initial energy normalized to  $E_n \equiv  q^2 B_a^2R^2/2m $, for three values of initial P and low $B_R$.  Below $P = 0.25$ orbits may be cyclotron or figure-8; above $P = 0.25$ orbits are betatron. A logarithmic drop in $\Delta f$ is seen for $P =  0.1$ and $0.2$ at an energy corresponding to the phase-space separatrix energy, at the transition of cyclotron into figure-8 orbits.  Little change in $\Delta f$ occurs for  betatron orbits.
 
In heating, the variance of energy, and therefore the maximum energy, $E_{max}$, will increase with time. Figure 3a) shows  $E_{max}$ attained by an initially figure-8 orbit as a function of time for four $B_R$ values and the same FRC parameters as in Fig. 2. $E_{max}$ displays saturation behavior quickly,  implying  phase coherence growing with increasing energy. The threshold for heating is  at $B_R \sim 3 $ G, above which $E_{max}$ grows $\propto B_R^{1.5}$.  Figure 3b) compares $E_{max}$ {\it vs} $B_R$ for figure-8 ($P=0.22$), betatron ($P=0.26$), and cyclotron ($P=0.19$)  orbits initiated in the $z=0$ subspace at the same radial position, $r/R=0.8131$, and the same energy, 1000 eV, for a simulation time of $\tau = 10^4$. Heating of figure-8 orbits occurs at lower $B_R$ than for cyclotron or betatron orbits. At $B_R \sim 20$ G, FFTs of energy for figure-8 ions initiated at higher energy, $E> 0.03 E_n \sim 30$ keV, in the $z=0$ subspace show sharp peaks and little further gain in energy.   The regular interaction with $RMF_0$ of these higher energy figure-8 orbits may be understood by the greater separation between resonances in phase-space\cite{Landsman:05, Landsmant:05} with increasing energy (or $\omega$).  

We have calculated the $RMF_o$-induced energy gain of a figure-8 orbit in a single half period of its motion as a first-order correction to the one-dimensional motion along $r$. For $P <0.25$, the shape of the effective potential, $V(r)$, is a 
double well\cite{Landsman:03}, corresponding to cyclotron orbits (in either well) inside the phase-space separatrix and to figure-8 orbits (moving across both wells) outside the phase-space separatrix.   The figure-8 orbit is approximated  by  motion in a symmetric double well:  $\rho \equiv \frac{r}{R}  = \rho_0 + a_1 cos\lbrack \omega \left(t-t_0\right) \rbrack +
a_2 cos \lbrack 3 \omega \left(t-t_0\right) \rbrack $, with $a_1/a_2 \sim 10$.  
 The amplitudes of oscillation,  $a_1$ and $a_2$, are determined  by the total energy.  The energy change from the interaction with the field is given by
${dH}/{dt} = q \vec \mathcal{E} \cdot \vec v = q \left(\mathcal{E}_r v_r + \mathcal{E}_\phi v_\phi \right)$,
with $c\vec \mathcal{E} = -{\partial \vec A}/{\partial t}$.
 After some algebra and integrating over a single half oscillation we get the energy gain from the interaction with the $\mathcal{E}_r$  and $ \mathcal{E}_\phi$ components of the RMF. The biggest energy change occurs for a resonance between
the $RMF_o$ and the Fourier components of the orbital motion, ${\it i.e.}$, $\omega_R/ \omega = s$,  where $s$ is an integer. The $ \mathcal{E}_r$- and $ \mathcal{E}_{\phi}$-induced radial and azimuthal portions of the energy change ($\triangle E$)  in a single half oscillation for an $s$ resonance   are:
\begin{equation}
\triangle E_r =  H_0 \left( \sum_{n=1}^{9,n \neq s} 
F_n(\omega)  cos(\Psi_0) + \left[\hat F_0 + F_s \right] sin(\Psi_0) \right)
\label{eq:delErres}
\end{equation}
\begin{equation}
\triangle E_\phi =  H_0 \left( \sum_{n=0}^{8,n \neq s} 
Q_n(\omega)  cos(\Psi_0) + Q_s(\omega) sin(\Psi_0) \right)
\label{eq:delEphires}
\end{equation}
where  $F_n(\omega) = C_n \left[ (-1)^{s+n} -1 \right] \frac{n}{s^2 - n^2}$, 
$\hat F_0= -\frac{C_0}{s} \left( (-1)^s -1\right)$, 
$F_s = \frac{\pi}{2} {C_s} $,  $Q_s(\omega) = \frac{\pi}{2} \left(\frac{\omega_{ci}}{\omega}\right) K_s$, $Q_n(\omega) = K_n \left[ (-1)^{s+n}-1 \right] \frac{s}{s^2 - n^2}  \frac{\omega_{ci}}{\omega} $,
$\Psi_0 = \phi - \omega_R t_0$,   $t_0$ is the initial time, $H_0 =   mk R^3  \omega_{ci}  \omega_RB_R/2B_a $, and the $C_n$ depend on $\rho_0$, $a_1$ and $a_2$. 
Since the total energy is  $H \sim \frac{1}{2} m (R \omega a_1)^2$ and
$|\Omega| \sim 1$, the relative fluctuations in energy during an oscillation are of order,   $max \triangle E_{odd}/{H} \sim O\left(10^2 s {B_R}/{B_a}\right)  $, and $max \triangle E_{even}/{H} \sim O\left(10 s^2 {B_R}/{B_a}\right).$ These predict significant energy gain for figure-8 orbits over a single oscillation, even for a relatively low amplitude RMF, $B_R / B_a \sim 10^{-3}$.   The energy gain for $s-even$ resonances has an $s^2$ dependence while $s-odd$ energy gain has a linear dependence on $s$.   Resonances with an odd value of $s$ show better heating than $s-even$ resonances, especially at lower values of $s$, where the ion energy is higher. 
Thus, the heating observed for figure-8 orbits at higher energies results primarily from an overlap of odd-$s$ resonances.    

Using the condition for exponential separation of trajectories\cite{Zaslavsky:98}, we now  determine the threshold for  the ergodicity of ion trajectories,   essential to convert energy gain to stochastic heating.  The change in energy over an oscillation   is used to map the dynamics:
$E_{j+1} = E_j + \triangle E(t_j)$;  $t_{j+1} = t_j + \frac{\pi}{\omega(E_{j+1})} $
where $t_j$ is the time of the start of successive ion oscillations at $\rho = \rho_{max} \equiv \rho_0 + a_1$ and $\triangle E(t_j)$ is $\triangle E_r(t_j) + \triangle E_{\phi}(t_j)$, with the substitutions $\Psi \to \Psi_j$,  $\Psi_j = \phi - \omega_R t_j$, and $ \Delta E \to \Delta E(t_j).$ The dynamics will be chaotic if  exponential separation of trajectories, {\it i.e.}, $K > 1$, occurs, where $K = max \left| \frac{d t_{j+1}}{dt_j} - 1 \right|$. In dimensionless variables,  $\tilde E = (m/b^2 R^2) E $ and $\tilde \omega = m \omega/b= 2 \omega/\omega_{ci}$, where $b = q B_a/2 c$.  $K$ for odd and even resonances are: 
\begin{equation}
K_{odd} \approx  8 \pi s  \left(\frac{1}{k R}\right) \left(\frac{B_R}{B_a}\right) \frac{d \tilde \omega (\tilde E)}{d \tilde  E}
\label{eq:Kodd}  
\end{equation}
\begin{equation}
K_{even} \approx  \frac{\pi}{2} s^2  \left(k R\right) 
\left(\frac{B_R}{B_a}\right) \frac{d \tilde \omega (\tilde E)}{d \tilde E}
\label{eq:Keven}  
\end{equation}
Based on these,  increasing the axial wavenumber, $k$, of the $RMF_o$ should lower the chaos threshold for $s-odd$ resonances  while raising the threshold for  $s-even$ resonances.  This is borne out by numerical simulation.

Fig. 3c) shows $d \tilde \omega (\tilde E)/d \tilde E$ ${\it vs.}$ $\tilde E$ for figure-8 orbits having $P=0.15$.   At energies very close to the separatrix,  $d \tilde \omega (\tilde E)/d \tilde E$ grows as  $(\tilde E-\tilde E_h)^{-5/6}$, where $\tilde E_h$ is the energy at the phase-space separatrix.   The large growth of $d \tilde \omega (\tilde E)/d \tilde E$ near the separatrix corresponds to a large increase in non-linearity  of the figure-8 orbital frequency.

The greatest  rate of stochastic heating is expected to occur for lower energy
figure-8 orbits where the values of $s$ and  $d \tilde \omega (\tilde
  E)/d \tilde E$ are higher.  As  $B_R$ is increased, the
  stochastic region above the phase-space separatix will broaden.  Close to the
  separatrix, even very low amplitudes of $B_R$ should produce chaotic
  orbits.  Eqns. 3) and 4), combined with Fig. 3c), 
can be used to estimate the relative amplitude of $B_R$ needed to produce stochasticity and  heating.  For example,  $s=3$ resonance occurs at $\tilde E \approx .0185$, corresponding to  
 $d \tilde \omega (\tilde E)/d \tilde E \approx 25$, see
    Fig. 3c).  Using Eqn. 3) and $k R
    \sim 1$, chaotic trajectories are to be expected for all $s-odd$
    resonances with $s \geq 3$ and $B_R/B_a \geq 5 \cdot 10^{-4}$, for
    $\Omega \sim 1$,  the assumption used in the
    derivation.  These findings approximately agree with the numerical findings.
Changing the value of a $P$  changes the scale of $\tilde E$, but does not have a substantial 
effect on the value of $d \tilde \omega (\tilde E)/d \tilde E$ at different resonances.  
Thus $P$ determines the energy range over which figure-8 orbits get
heated, with greater energy range for lower values of $P$, while not affecting
the approximate structure of phase space.
In Fig. 3c), all  $s>4$ resonances are located to the left of $\tilde E \approx 0.013$, hence occur over the interval  $\delta \tilde E \sim 0.003$.
This leads to much greater chaos closer to the phase-space separatrix where
the closely spaced resonances overlap.  Thus, lower-energy figure-8
orbits are more chaotic and much better heated by the $RMF_o$ than the
higher energy ones.  Fig. 3d) shows this effect for two values of P:  0.17 and 0.2.  Figure-8 orbits are not further heated once their energy reaches (or initially exceeds) the curved line appropriate for each P value.  The simplifications on which Eqns. 3) and 4) are based become less accurate at $B_R/B_a > 0.001$.

Among the clearest differences between these results for the FRC and those reported  for the tokamak are: 1) The non-linearities for the FRC arise from the double potential well and field gradient and their direct effects on the particle orbit. Those in the tokamak arise from trapping in the wave field -- hence require a stronger wave field -- and resonance between the cyclotron motion and the wave field, resulting in a single large resonance and large first-order islands.  In contrast, close spacing in phase space between resonances of a figure-8 orbit leads to an  overlap between resonances and the observed  stochastic heating for figure-8 orbits in FRCs.
The importance of the time-varying field, $\mathcal{E}$, in the tokamak analysis is to create a small nonlinearity in this 2-D system (of the order of $\mathcal{E}/B$)  which leads to resonances between the two degrees-of-freedom, not between the $\mathcal{E}$ field and the ion trajectory. 2) In the tokamak, heating occurs at $10 \times$ higher values of  $\Omega$ (over 20 {\it vs} 1 in the FRC) and lower values of $\omega_R/kv_{i,thermal}$ ($\sim 1$ {\it vs} 10). 3) The threshold for heating in the FRC is lower by the factor $ s  d \tilde \omega (\tilde E)/d \tilde E$, through which the effect of the FRC's double effective-potential well is clear.

In summary, the energy gain  in an orbital period due to $RMF_o$ was calculated for a figure-8 orbit in an FRC. Resonances of $\omega_R$ with $\omega$ produce significant energy gain.
Odd-$s$ resonances  more effectively heat  for
high energy (lower $s$) figure-8 orbits.  The energy gain in a
oscillation was used to map the dynamics and a criterion for the
exponential separation of trajectories was used to find the threshold for chaotic orbits. $K$, the measure of the rate of trajectory separation,  increases with $B_R$.   
At higher energies, the orbits are less chaotic due to both a lower
value of $s$ and, more importantly, to a  decreased nonlinearity  reducing $d \tilde \omega (\tilde E)/d \tilde E$.

 
  \begin{figure}[htb] {\begin{center}
\scalebox{.7}[.7]
{\includegraphics{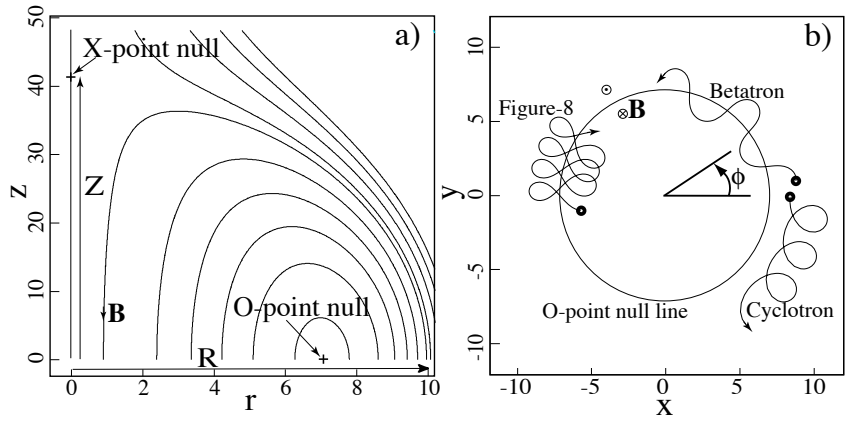}} \end{center} }
\vspace {-0.9cm}
\caption{\small {  a) Shape of FRC magnetic field in z-r plane. $\kappa = 4.1$. b) Shapes of typical cyclotron, betatron and figure-8 orbits in $z=0$ plane.  } }
\end{figure} 
 
 \begin{figure}[htb] {\begin{center}
\scalebox{.7}[.7]
{\includegraphics{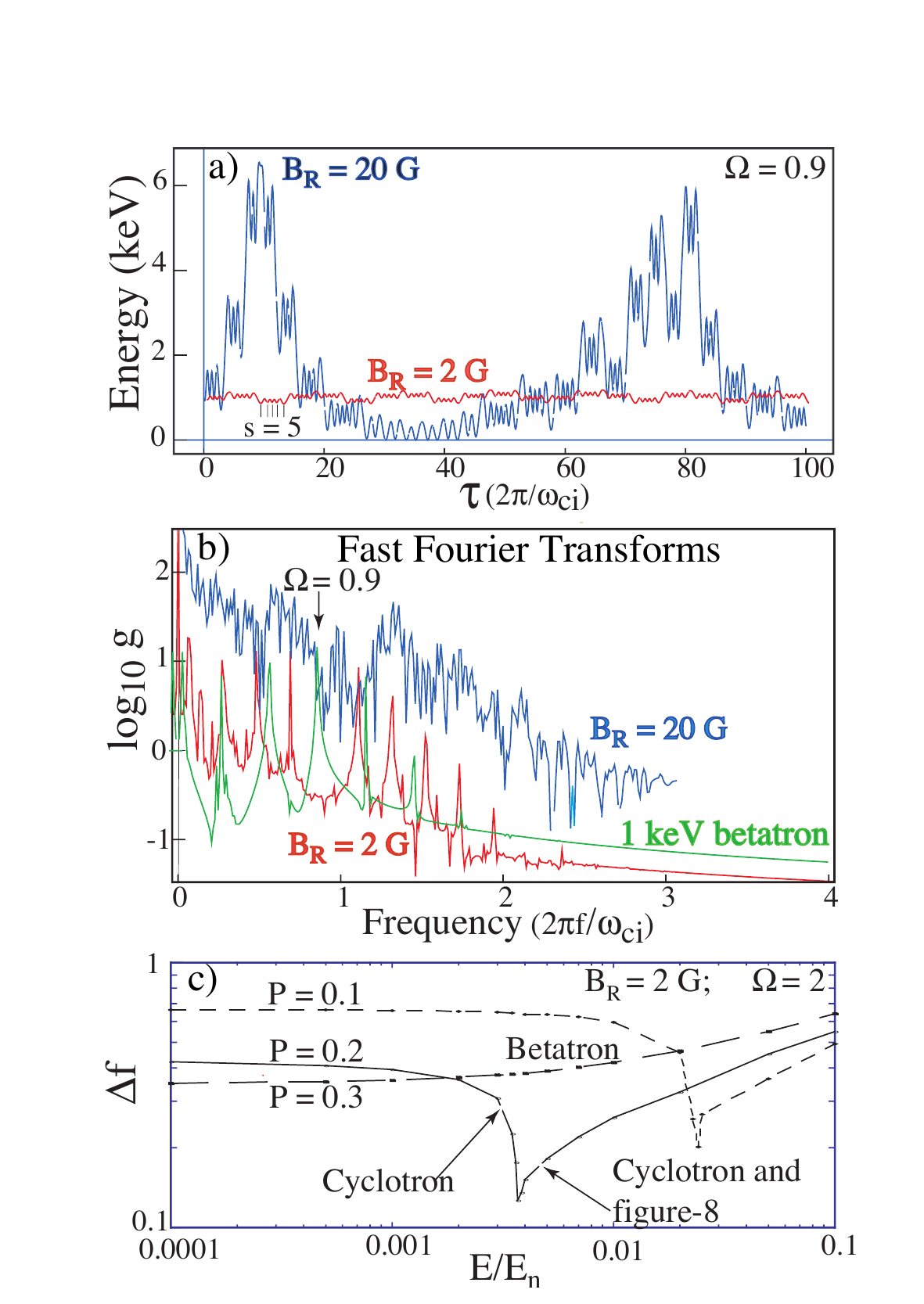}} \end{center} }
\vspace {-0.9cm}
\caption{\small { Results of numerical simulation of ion trajectories in an FRC with $RMF_o$. a) Figure-8 ion energy {\it versus} time for two values of $B_R$. b) FFTs of ion energy for the two cases shown in a) and also for a betatron orbit.  c) Separation between resonances, $\Delta f$, {\it versus} normalized initial  energy for  P = 0.1, 0.2 and 0.3, at low $B_R$.  } }
\end{figure} 

 \begin{figure}[htb] {\begin{center}
\scalebox{.8}[.8]
{\includegraphics{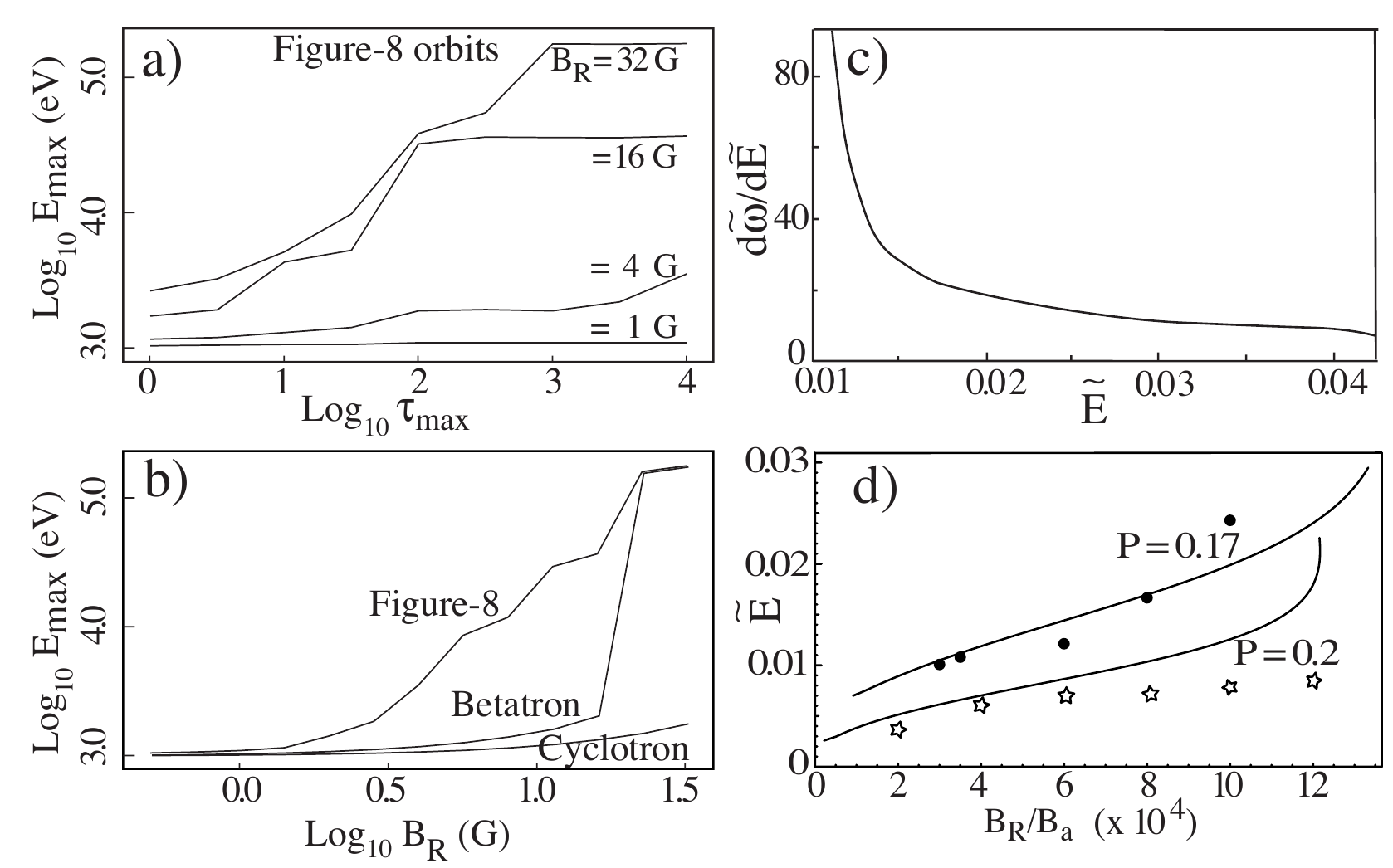}} \end{center} }
\vspace {-0.1cm}
\caption{\small { a) Maximum energy attained {\it versus} time for four values of $B_R$.  b)  Maximum energy attained by $\tau = 10^4$ {\it versus} $B_R$ by three types of orbits. c) $d \tilde \omega (\tilde E)/d \tilde E$ ${\it versus}$ $\tilde E$ for figure-8 orbits.  $P=0.15$.  d) Curves: Threshold $\tilde E$  for saturation of stochastic heating {\it versus}  $B_R/B_a$, based on Eqns 3) and 4):  $P=0.17$ (upper curve) and $P=0.2$ (lower curve);  Data points, dots and stars, from RMF-code.} }
\end{figure}

\end{document}